\documentclass[aps,prb,preprint,superscriptaddress,showpacs]{revtex4}
\usepackage{graphicx}
\begin{document}
\title{Field-induced transition between magnetically disordered and
ordered phases in underdoped La$_{2-x}$Sr$_x$CuO$_{4}$}
\author{B. Khaykovich}
\affiliation{Department of Physics, Massachusetts Institute of
Technology, Cambridge, Massachusetts 02139}
\author{S. Wakimoto}
\affiliation{Japan Atomic Energy Research Institute, Advanced
Science Research Center, Tokai, Ibaraki 319-1195, Japan}
\author{R. J. Birgeneau}
\affiliation{Department of Physics, University of California at
Berkeley, Berkeley, California 94720}
\author{M. A. Kastner}
\author{Y. S. Lee}
\affiliation{Department of Physics, Massachusetts Institute of
Technology, Cambridge, Massachusetts 02139}
\author{P. Smeibidl}
\author{P. Vorderwisch}
\affiliation{BENSC, Hahn-Meitner Institute, D-14109 Berlin,
     Germany}
\author{K. Yamada}
\affiliation{Institute for Materials Research (IMR), Tohoku
University Katahira, Sendai 980-8577, Japan }
\date{\today}
\begin{abstract}
We report the observation of a magnetic-field-induced transition
between magnetically disordered and ordered phases in slightly
under-doped La$_{2-x}$Sr$_{x}$CuO$_4$ with $x=0.144$. Static
incommensurate spin-density-wave order is induced above a critical
field of about 3 T, as measured by elastic neutron scattering. Our
results allow us to constrain the location of a quantum critical
point on the phase diagram.
\end{abstract}
\maketitle

The interplay between magnetic order and superconducting order is
one of the most fascinating and important subjects in the study of
high-$T_c$ cuprate
superconductors.\cite{kastner-rmp98,Millis-Orenstein-review,%
kivelson-RMP,sachdev-rmp03} Magnetic order or fluctuations are
found in almost every member of this large family of materials, in
both hole-doped and electron-doped superconductors, and in the
entire doping range studied, from undoped antiferromagnetic
insulators to the overdoped superconductors. At first, the
magnetism in different cuprates appears to have different
manifestations: incommensurate static or dynamic
spin-density-waves in
La$_{2-x}$Sr$_{x}$CuO$_4$,\cite{kastner-rmp98} a resonance peak at
the antiferromagnetic
zone center in YBa$_2$Cu${_3}$O$_{6+x}$,%
\cite{YBCO-resonance-Rossat,YBCO-resonance-Mook,YBCO-resonance-Keimer}
and Bi$_2$Sr$_2$CaCu$_2$O$_8$,\cite{BSCCO-resonance-Keimer} and 3D
antiferromagnetic order in electron-doped Nd$_{2-x}$Ce$_x$CuO$_4$
and Pr$_{1-x}$LaCe$_x$CuO$_4$.\cite{Uefuji-PhysicaC01,fujita-03}
However, recent experiments suggest a more unified picture, at
least for the hole-doped materials. It is found that
characteristic features of both spin-density correlations and the
resonance exist in
La$_{2-x}$Sr$_{x}$CuO$_4$,\cite{Tranquada-LSCOinField}
La$_{2-x}$Ba$_{x}$CuO$_4$,\cite{Tranquada-LBCOresonance} and
YBa$_2$Cu${_3}$O$_{6+x}$.\cite{mook-nature98,stock-prb04,hayden-04}
In addition, high-energy neutron scattering studies show that the
magnetic fluctuations have similar dispersion in
La$_{2-x}$Ba$_{x}$CuO$_4$,\cite{Tranquada-LBCOresonance}
optimally-doped La$_{2-x}$Sr$_{x}$CuO$_4$ \cite{christensen-04}
and YBa$_2$Cu${_3}$O$_{6+x}$.\cite{arai-prl99,bourges-science2000}
The importance of these results is that La$_{2-x}$Sr$_{x}$CuO$_4$
and YBa$_2$Cu${_3}$O$_{6+x}$ have very different superconducting
$T_c$'s, crystal structures, and other properties. Therefore, the
This universality of the magnetic properties of doped cuprates
implies a fundamental role that spin fluctuations play in the
physics, and a complete understanding of the magnetism has become
even more important.

Many cuprates have a tendency to develop stripe-like spin
correlations. In doped La$_{2}$CuO$_4$, when the doping is below
optimal for the superconductivity, the stripe-like correlations
become static, forming incommensurate spin-density
waves.\cite{tranquada-nature95} For optimal and higher doping the
static order disappears, but dynamic correlations persist up to
the superconductor-normal-metal boundary.\cite{waki-overdoped} The
disappearance of the stripe-like correlations at the
superconductor-normal-metal boundary suggests that dynamic stripes
are necessary for the superconductivity, as some theories have
predicted.\cite{EmeryKivelsonZachar} However, \emph{static}
magnetic order clearly competes with the superconducting
order.\cite{tranquada-nature95} The nature of this competition has
been studied by recent magnetic field experiments on
La$_2$CuO$_{4+y}$,\cite{khaykovich-prb02,khaykovich-prb03} and
La$_{2-x}$Sr$_x$CuO$_4$.\cite{katano-prb00,lake-nature02} These
experiments show that the magnetic moment associated with static
long-range SDW order increases when the superconductivity is
weakened by the applied magnetic field. Note that in
electron-doped cuprate superconductors, like
Nd$_{2-x}$Ce$_x$CuO$_4$ and Pr$_{1-x}$LaCe$_x$CuO$_4$, static
magnetism also coexists with the superconductivity. In this case
too, it appears that the ordered magnetic moment increases with
applied magnetic field,\cite{fujita-03,matsuura-prb04} but it is
not clear whether the mechanism responsible for this increase is
the same as in the hole-doped cuprates.\cite{Mang-04}

Magnetic correlations in La$_{2-x}$Sr$_x$CuO$_4$ reveal themselves
in neutron scattering measurements as a quartet of incommensurate
peaks around the antiferromagnetic zone center (at reciprocal
lattice positions $(1\pm\delta, \pm \delta)$ in orthorhombic
notation, where $\delta \simeq 0.125$ for $x\geq 0.125$). In
La$_{2-x}$Sr$_x$CuO$_4$, previous studies have shown that static
magnetic order exists for $x \leq 0.13$, while for larger $x$
there are gapped spin excitations at the same positions in
reciprocal space; gap values of 4 to 8 meV are measured for $x
\geq 0.15$ depending on the doping
level.\cite{yamada-prl95,LeeYamada-prb03} Previous neutron
scattering experiments in magnetic fields have revealed either an
enhancement of the static SDW order in underdoped
La$_{2-x}$Sr$_x$CuO$_4$ and La$_2$CuO$_{4+y}$ superconductors
\cite{katano-prb00,lake-nature02,khaykovich-prb02,khaykovich-prb03}
or changes in the dynamic spin susceptibility in optimally doped
or slightly overdoped
La$_{2-x}$Sr$_x$CuO$_4$.\cite{lake-science01,gilardi,Tranquada-LSCOinField}
In the former case, the intensity of the magnetic Bragg peaks
increases with magnetic field accompanying the suppression of the
superconducting order parameter. In the latter case, the magnetic
field enhances the spin susceptibility at energies below the gap.

\begin{figure}
\centering
\includegraphics[width=4.5in]{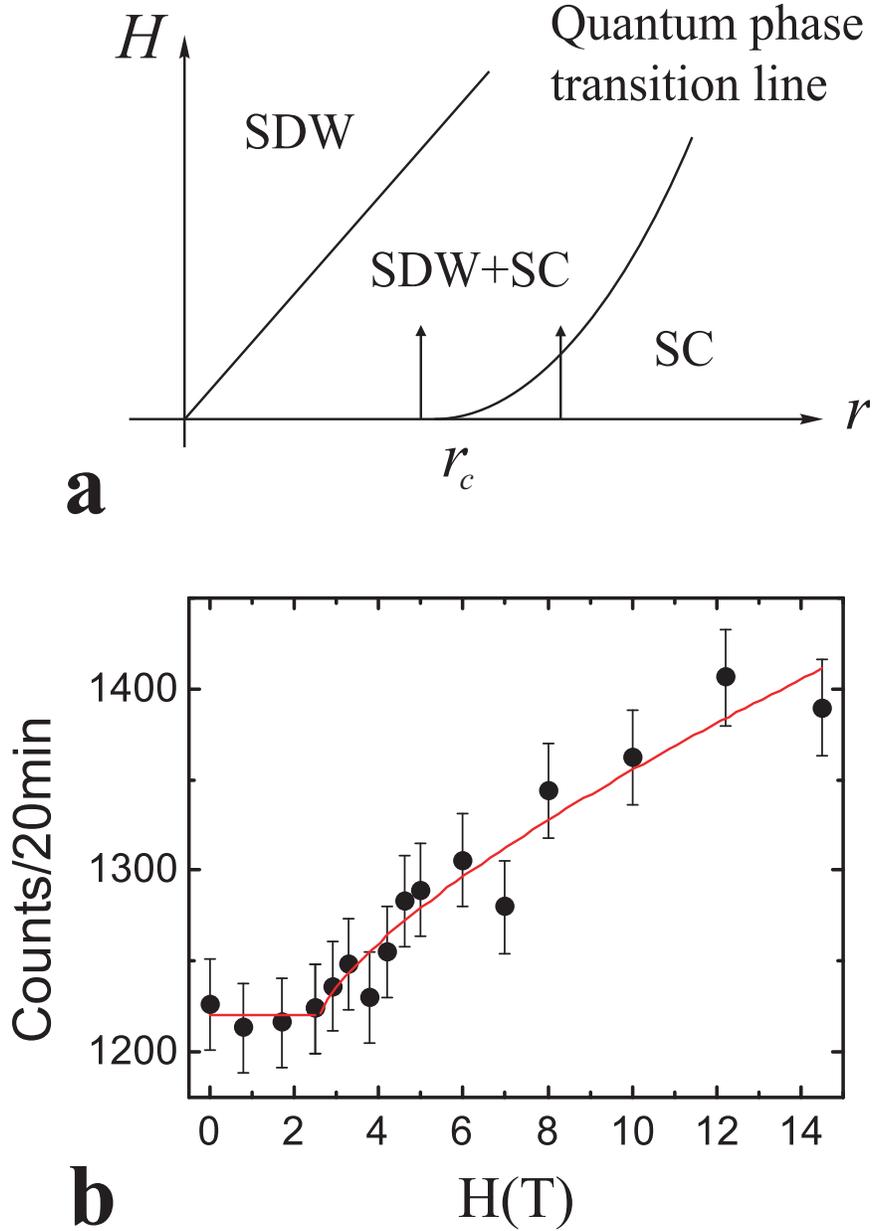} 
\vspace{0in}%
\caption{\label{FieldDep}(a) A fragment of the theoretical phase
    diagram, adopted from Refs. \onlinecite{demler-prl01,sachdev-rmp03}.
    The vertical axis is the magnetic field and the horizontal axis is
    the coupling strength between superconductivity and magnetic order.
    (b) Field dependence of the magnetic Bragg peak corresponding to the
    incommensurate SDW peak at Q = (1.125,0.125,0). Every point is measured
    after field cooling at T = 1.5 K. The data are fitted to
    $I=I_0+A|H-H_c|^{2\beta}$ above $H_c$ as explained in the text.
    Spectrometer configuration: 45-60-Be-S-Be-60-open; cold Be filters were
    used before and after the sample to eliminate contamination from high
    energy neutrons; E = 4 meV. }
\end{figure}

As a result of these experiments, a theoretical phase diagram of
doped La$_2$CuO$_{4}$ superconductors has been proposed for T = 0
K.\cite{demler-prl01,kivelson-prb02} Fig. \ref{FieldDep}a shows a
fragment of this phase diagram, adopted from Refs.
\onlinecite{sachdev-rmp03,demler-prl01}. The horizontal axis $r$
is a measure of the repulsive coupling between the superconducting
and magnetic order parameters; it is assumed to be approximately
proportional to doping. The vertical axis is the magnetic field.
There exist three distinct phases: (i) the spin-density-wave (SDW)
phase at high field, presumably above the upper critical field,
where superconductivity is destroyed by the field, (ii) the
superconducting (SC) phase at high doping and small fields, and
(iii) the intermediate ``SC+SDW'' phase, where the SC and SDW
order parameters coexist. The boundary between the SC and SC+SDW
phases is a line of quantum phase transitions. The left vertical
arrow shows the approximate trajectory through the phase diagram
for the earlier experiments on underdoped La$_{2-x}$Sr$_x$CuO$_4$
and La$_2$CuO$_{4+y}$.
\cite{katano-prb00,lake-nature02,khaykovich-prb02,khaykovich-prb03}
These materials are in the phase where the magnetic and
superconducting order parameters co-exist and compete
microscopically. This phase is transformed into a SC magnetically
disordered phase with increased doping. In the latter phase, the
spin susceptibility increases with magnetic field at low energies
because the applied field brings the system closer to the magnetic
ordering transition while suppressing the superconducting order.
The phase diagram predicts that the spin-ordering transition can
be achieved either by reducing the doping or applying a strong
enough magnetic field. However, substitutional doping affects both
the superconducting and magnetic order parameters, as well as the
degree of disorder in the samples. In this sense, achieving the
phase transition by applying a magnetic field would be the most
straightforward measurement of the phase transition.

The experiments presented here demonstrate directly that the
spin-ordering transition can be induced by a magnetic field in
slightly underdoped La$_{2-x}$Sr$_{x}$CuO$_4$ with nominal
$x=0.144$. Fig. \ref{FieldDep}b shows the field dependence of one
of the incommensurate magnetic Bragg peaks for this material. At
zero applied field, no elastic signal is found, as expected for
this doping level. Static long-range magnetic order appears above
approximately 3 T. This result is direct evidence of the quantum
phase transition between magnetically-ordered and disordered
superconducting states in La$_{2-x}$Sr$_x$CuO$_4$.


The single crystal of La$_{2-x}$Sr$_x$CuO$_4$ with nominal Sr
concentration $x= 0.144$ was grown by the travelling solvent
floating zone technique with subsequent annealing in oxygen
atmosphere at 900 $^{\circ}$C for 24 hours followed by furnace
cooling in oxygen. We estimate the present sample doping $x=
0.144\pm 0.005$. The preparation conditions are essentially
identical to other samples of La$_{2-x}$Sr$_x$CuO$_4$ ($x=0.1$,
0.12, 0.13, 0.15, etc.) used to study the SDW order and
fluctuations (see refs. \onlinecite{yamada-prl95,katano-prb00} and
refs. therein), which assures consistency in Sr and oxygen
contents. Magnetic susceptibility of a small piece cut from the
neutron scattering sample showed a superconducting transition with
$T_c$ (onset) = 37 K, centered at 35 K. The sample of 8 g was cut
into two equal pieces, which were co-mounted in order to fit
inside the split-coil 14.5 T magnet; the mosaic of the assembly
was $0.5^{\circ}$ (FWHM). Elastic neutron scattering experiments
in the presence of a high magnetic field were performed at the
Hahn-Meitner Institute, utilizing the FLEX cold-neutron
triple-axis spectrometer with incident neutron energy of 4 meV.

Fig. \ref{FieldDep}b shows the field dependence of the intensity
of one of the SDW peaks in the $x=0.144$ sample. Similar behavior
is found for the other incommensurate peaks in this sample. The
intensity is fitted according to a power law
$I(H)=I_0+A(H-H_c)^{2\beta}$, as expected near a second-order
phase transition. Here $I_0$ is a constant background,
$H_c=(2.7\pm0.8)$ T and $\beta = (0.36\pm0.10)$ are fitting
parameters. The fit is shown as a solid curve in Fig.
\ref{FieldDep}b. The current statistical error bars do not allow
us to draw conclusions about the universality class of the
transition.

\begin{figure}
\centering
\includegraphics[width=6.5in]{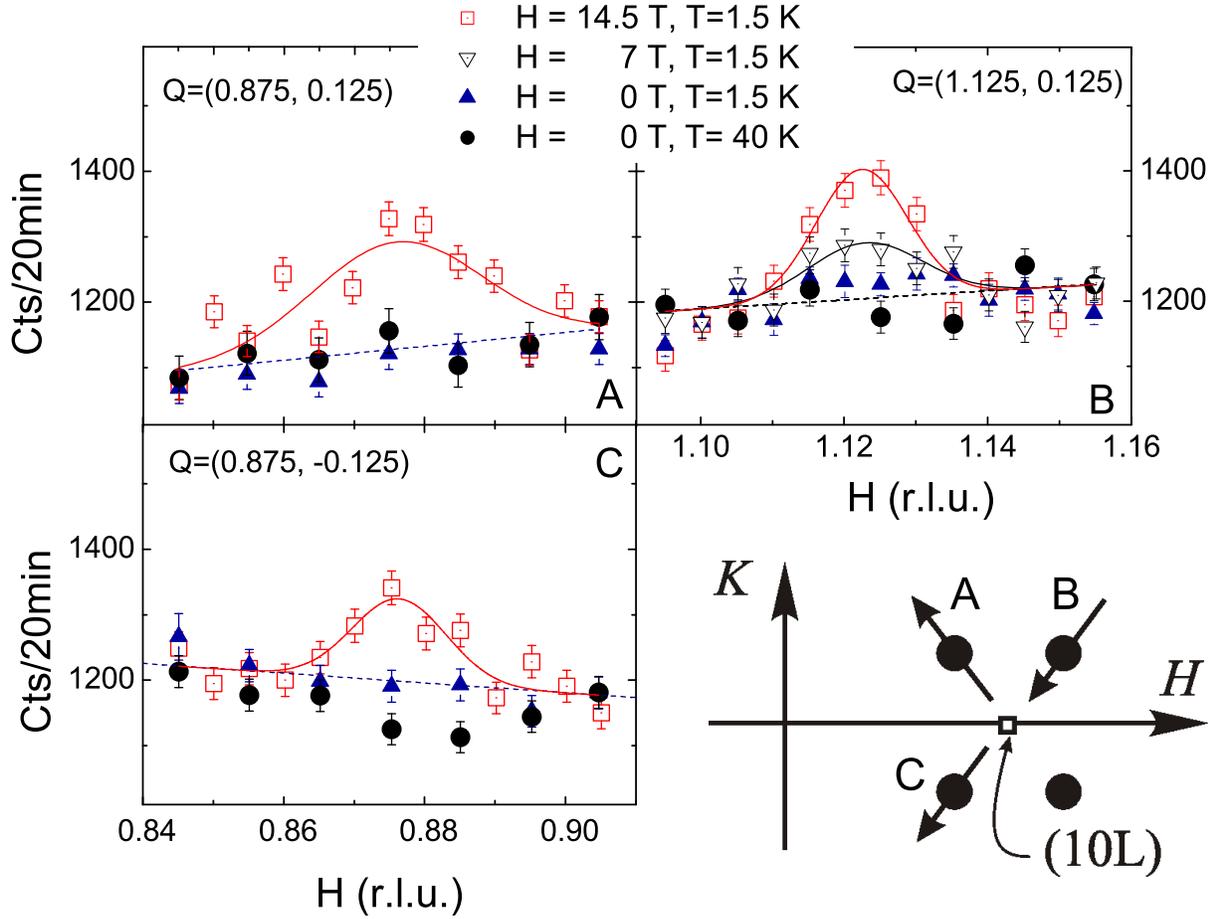} 
\vspace{0in}%
\caption{\label{144Sample} Incommensurate SDW peaks in the
    $x=0.144$ sample. The peaks are shown at three
    different positions around the AFM zone center at different
    magnetic fields for $T = 1.5$ K and for $H = 0$ T at 40 K,
    which is above $T_c = 37$ K and above the highest temperature
    at which the SDW order is formed in La$_{2-x}$Sr$_{x}$CuO$_4$.
    The peak positions and scans directions are
    shown schematically in the inset. Magnetic field scans were
    performed after field-cooling. The zero-field background data are
    fitted by a linear function and the field-induced magnetic Bragg
    peaks are fitted by a Gaussian above this linear background.}
\end{figure}

Fig. \ref{144Sample} shows three out of four incommensurate SDW
peaks. Comparing the scans at $H = 0$ T for $T=40$ K and $1.5$ K
shows that the scans in zero field do not change above and below
the superconducting $T_c$ within the errors. The absence of a peak
places an upper bound on the zero-field ordered magnetic moment of
0.02$\mu_B$, assuming a resolution-limited correlation length. The
peaks in Fig. \ref{144Sample} B and C are resolution limited at
14.5 T, which places a lower limit on the magnetic correlation
length of $120 \AA$ (the data in Fig. \ref{144Sample} A is too
noisy to allow for a reliable fit). The ordered moment at 14.5 T
is $(0.06 \pm 0.02)\mu_B $.\cite{mag-moment} Fig. \ref{144Sample}B
also shows the peak at $H = 7$ T, where the intensity is
approximately half of that at $H = 14.5$ T.


The behavior of the $x=0.144$ sample is qualitatively and
quantitatively different from that of previously studied samples
of La$_{2-x}$Sr$_x$CuO$_4$ (\emph{x} = 0.1 and 0.12) and
La$_2$CuO$_{4+y}$, where an enhancement of an \emph{existing}
static SDW order was
found.\cite{katano-prb00,lake-nature02,khaykovich-prb02,khaykovich-prb03}
As mentioned above, the scans in Fig. \ref{144Sample} suggest that
for $x=0.144$ the magnetic moment is zero when $H = 0$ T. In
addition, the field dependence shown in Fig. \ref{FieldDep}b is
consistent with the absence of SDW order below $\sim 2.7$ T. In
the previous results for La$_{1.9}$Sr$_{0.1}$CuO$_4$, the
field-induced signal at 2.5 Tesla is about 35\% of the signal
induced at 14 T.\cite{lake-nature02} If a comparable increase
existed for our sample, the signal at 2.5 Tesla should be $\sim
1280$ counts/20 min. As seen in Fig. \ref{FieldDep}b, such a
signal is not observed and would be clearly outside of the error
bars. In addition, the field dependence for
La$_{1.9}$Sr$_{0.1}$CuO$_4$ is fit by a line which has an infinite
slope in limit of $H \rightarrow 0$. Again, our data do not
support this and are qualitatively different. The doping of $x =
0.144$ was chosen to be close to $x = 0.15$ for which it is known
that the ground state has complete spin-gap.

It is useful now to refer to the phase diagram in the Fig.
\ref{FieldDep}a. The left vertical arrow corresponds to the
$x=0.10, 0.12$ and La$_2$CuO$_{4+y}$ samples, which are in the
``SC+SDW'' phase at zero field.
\cite{katano-prb00,lake-nature02,khaykovich-prb02,khaykovich-prb03}
By contrast, the $x=0.144$ sample is in the SC phase at $H=0$ T
and crosses into the SC+SDW phase as the field increases. This
behavior corresponds to the right vertical arrow in the phase
diagram. Note that the high-field SDW state is an actual
thermodynamic phase since we observe long-range SDW order above
the transition. Given that a static, long-range ordered SDW phase
is found at $H = 0$ in La$_{2-x}$Sr$_x$CuO$_4$ for $x=0.12$,
\cite{kimura-prb99} but not in our sample with $x=0.144 \pm
0.005$, we conclude that the quantum critical point at $H=0$ T
lies between $x=0.12$ and $x=0.14$. We will discuss this further
below.

Most models for the enhancement of SDW order or fluctuations by
the magnetic field assume a competition between the
superconductivity and magnetic order, although different models
for this competition have been
proposed.\cite{zaanen-prb89,zhang-science97,demler-prl01,%
kivelson-prb02,DHLee-prl02} Our experimental results, combined
with the earlier experiments
\cite{katano-prb00,lake-nature02,khaykovich-prb02,khaykovich-prb03}
favor the model proposed in Ref. \onlinecite{demler-prl01}. This
model assumes a microscopic competition between SDW and SC order.
This is a Ginzburg-Landau model with a repulsive coupling between
the superconducting and magnetic order parameters near the
magnetic ordering phase transition. The magnetic field penetrates
type-II superconductors in the form of magnetic flux lines and the
superconducting order parameter is suppressed to zero inside the
vortex cores and recovers to its zero-field value over a large
length scale away from the cores. This suppression of the
superconducting order parameter leads to an enhancement in the
competing magnetic order far away from the vortex cores and
ensures that the magnetic correlation length spans many
inter-vortex distances.

This leads us to propose a possible scenario for the field-induced
transition. It is well known that in the SC phase the spins
fluctuate at various frequencies, which may vary depending on
local disorder or doping variations. These 2-dimensional spin
fluctuations at $Q =(1\pm\delta,\pm\delta)$ around the
antiferromagnetic zone center are well characterized by inelastic
neutron scattering.\cite{kastner-rmp98,aeppli-science97} When a
magnetic field is applied, the superconductivity is completely
suppressed within the vortex cores, and partially suppressed even
at large distances from the vortices. Because of the competition
between SC and SDW order the magnetic fluctuations are enhanced
where SC is suppressed, and their spectral weight distribution is
modified such that lower frequency fluctuations become
stronger.\cite{lake-science01,gilardi,Tranquada-LSCOinField} When
the field exceeds a critical value, the regions of slowest spin
fluctuations become large enough that long-range magnetic order
can develop in the form of static incommensurate SDW order at the
same $Q =(1\pm\delta,\pm\delta)$. When the field is increased
further, the magnetic order parameter is stabilized at the expense
of the superconducting order. It is expected that away from the
critical regime near $H_{c}(x)$ (and well below $H_{c2}$ where the
superconductivity is completely destroyed), the elastic neutron
scattering intensity should follow an $(H/H_{c2})\ln(H_{c2}/H)$
dependence,\cite{DS-private} similar to the samples where the SDW
order exists at zero
field.\cite{lake-nature02,khaykovich-prb02,khaykovich-prb03}
However, our experiments are close enough to the phase transition
that they are dominated by the critical fluctuations. The model of
Ref. \onlinecite{demler-prl01} predicts a linear behavior of the
intensity with the field, which is not inconsistent with the data
of Fig. \ref{FieldDep}b.\cite{DS-private}

In the La$_{2-x}$Sr$_x$CuO$_4$ phase diagram, static magnetic
correlations are observed for $x \leq 0.13$. However, the static
correlation length is short-ranged for $x=0.13$ ($\xi \simeq 88
\AA$), \cite{matsushita-99} whereas it diverges for $x \simeq
0.12$ ($\xi > 200 \AA$).\cite{kimura-prb99} Therefore, the $H = 0$
quantum critical point separating the SC+SDW and SC phases is
close to $x=0.125$. The short-range SDW peaks observed for larger
$x$ would then correspond to critical scattering. Indeed, for a
sample with $x=0.14$, we have observed short-range SDW order in
zero-field and a transition to long-range order with increasing
field, \cite{0.14_sample} consistent with this picture. For
concentrations near but below $x=0.125$, the SDW peaks are
resolution-limited, and long-range order is observed. However, as
$x$ is decreased further, the static correlation length at low
temperatures is again measurable and finite.\cite{fujita-02} This
behavior is consistent with an incommensurate system with
structural (i.e., random field) disorder. As the system becomes
more incommensurate, the random fields become more important. The
QCP appears near $x=0.125$ because that is the concentration at
which the system is closest to being commensurate. Note that
random fields are much less important for commensurate systems.
Similar behavior has been recently discussed in relation to an
avoided critical point on the phase diagram by Kivelson and
coworkers.\cite{kivelson-prb02}

Finally, we note a possible connection between the present results
and the spin gap phenomenon, which is observed in
La$_{2-x}$Sr$_x$CuO$_4$ for $x \geq
0.15$.\cite{matsuda-prb94,LeeYamada-prb03} One theoretical model
associates the spin gap with the proximity to a quantum phase
transition between the superconducting phase with no magnetic
order and a SC+SDW phase, where SC and SDW order parameters
coexist.\cite{kivelson-RMP} The spin fluctuations in the
disordered phase are expected to be gapped near a transition to
the magnetically-ordered phase because of the critical slowing
down of the quantum fluctuations. The gap energy is $E_{gap} \sim
\hbar/\tau$, where $\tau$ is the characteristic time scale of spin
fluctuations. This behavior is in contrast to a classical phase
transition, where the Heisenberg uncertainty relation does not
apply and the spin-gap is not expected to form in the disordered
phase. Earlier indirect evidence for a presence of quantum phase
transition has come from experiments on La$_{2-x}$Sr$_x$CuO$_4$,
$x \simeq 0.14$, which showed a diverging correlation length of
the spin fluctuations, but do not identify the two phases that are
separated by this transition.\cite{aeppli-science97} Inelastic
neutron scattering measurements on our $x=0.144$ sample would be
of great interest to explore the relationship between the spin-gap
in La$_{2-x}$Sr$_{x}$CuO$_4$ and the spin-ordering phase
transition reported here.

In conclusion, we have observed that in slightly underdoped
La$_{2-x}$Sr$_x$CuO$_4$ there is a magnetic-field-induced
transition between the magnetically disordered state and the
magnetically order state with incommensurate SDW order coexisting
with the superconductivity. In addition, since long-range SDW
order can be induced by an applied field, it appears that the SDW
phase is not the result of chemical inhomogeneities, but rather is
a true thermodynamic phase. Our results are consistent with a
picture in which the $H = 0$ quantum-critical point lies close to
the doping level of $x = 0.125$. Further experiments, especially
measurements of the instantaneous correlation length by neutron
scattering, would be most interesting to study the critical
behavior near this point.

\begin{acknowledgments}
We thank E. Demler and S. Sachdev for useful discussions. The work
at MIT was supported by National Science Foundation under award
number DMR-0239377. The work in Tohoku University was supported by
the Japanese Ministry of Education, Culture, Sports, Science and
Technology, Grant-in-Aid for Scientific Research.
\end{acknowledgments}

\begin{thebibliography}{44}
\bibitem[{mag()}]{mag-moment}
\bibinfo{note}{Ordered moment was calculated by comparing the integrated
  intensities of the SDW peaks with those of the oxygen-doped La$_2$CuO$_{4+y}$
  sample measured on the same spectrometer at the same
  configuration.\cite{khaykovich-prb03} Magnetic moment in this
  La$_2$CuO$_{4+y}$ sample has been characterized extensively with consistent
  result of $0.15 \pm 0.05 \mu_B$ (ref. \onlinecite{khaykovich-prb03} and
  references therein).}
\end{thebibliography}

\end{document}